\documentclass{elsart}

\usepackage{graphicx}

\begin{document}

\newcommand{\snn}{\sqrt{s_{NN}}}
\newcommand{\seff}{\sqrt{s_{\rm eff}}}
\newcommand{\s}{\sqrt{s}}
\newcommand{\pp}{pp}
\newcommand{\aaa}{A+A}
\newcommand{\pbarp}{\overline{p}p}
\newcommand{\qbarq}{\overline{q}q}
\newcommand{\epem}{e^+e^-}

\newcommand{\nhit}{N_{hit}}
\newcommand{\npp}{n_{pp}}
\newcommand{\nch}{N_{ch}}
\newcommand{\nchee}{N^{\epem}_{ch}}
\newcommand{\nchaa}{N^{\aaa}_{ch}}
\newcommand{\np}{N_{part}}
\newcommand{\ns}{N_{spec}}
\newcommand{\ntot}{\langle\nch\rangle}
\newcommand{\avenp}{\langle\np\rangle}
\newcommand{\npB}{N_{part}^B}
\newcommand{\nc}{N_{coll}}
\newcommand{\avenc}{\langle\nc\rangle}
\newcommand{\halfx}{\frac{1}{2}}
\newcommand{\halfnp}{\langle\np/2\rangle}
\newcommand{\etap}{\eta^{\prime}}
\newcommand{\as}{\alpha_{s}(s)}
\newcommand{\etazero}{\eta = 0}
\newcommand{\etaone}{|\eta| < 1}
\newcommand{\dndeta}{d\nch/d\eta}
\newcommand{\dndetazero}{\dndeta|_{\etazero}}
\newcommand{\dndetaone}{\dndeta|_{\etaone}}
\newcommand{\dndetanp}{\dndeta / \halfnp}
\newcommand{\dndetaonp}{\dndeta / \np}
\newcommand{\dndetazeronp}{\dndetazero / \halfnp}
\newcommand{\dndetaonenp}{\dndetaone / \halfnp}
\newcommand{\ratio}{\ntot/\halfnp}
\newcommand{\nee}{N_{ee}}
\newcommand{\nhh}{N_{hh}}
\newcommand{\nubar}{\overline{\nu}}

\begin{frontmatter}

\title{The Origin of the Difference between Multiplicities in $\epem$ 
Annihilation and Heavy Ion Collisions.}

\author{J.~Cleymans$^{1}$, M.~Stankiewicz$^{1}$, P. Steinberg$^{1,2}$, S.~Wheaton$^{1}$}
\address{$^{1}$Department  of  Physics and UCT-CERN Research Centre\\University of Cape Town, Rondebosch 7701, South Africa}
\address{$^{2}$Chemistry Department\\Brookhaven National Laboratory, Upton, NY 11973-5000, USA}
\date{\today}

\begin{abstract}
Multiplicities in $\epem$ annihilation and relativistic heavy ion collisions
show remarkable similarities at high energies.  A thermal-statistical
 model is proposed to explain 
the differences which occur mainly at low beam energies.
Two different
calculations are performed, one using an approximate thermodynamic
relationship, the other using a full thermal model code.
The results are in qualitative agreement, suggesting that
the interplay of baryon density and temperature tends to
systematically suppress the total multiplicity at lower
beam energies.
\end{abstract}
\end{frontmatter}

Recently, the PHOBOS experiment at RHIC has shown results on the
total charged particle multiplicity produced in heavy ion
collisions over a wide range of collision energies as well as
the collision centrality, characterized by the number of
participating nucleons~\cite{Back:2003xk}. 
Three surprising connections emerge
from comparison of this data with particle multiplicities 
measured in elementary collisions ($\pp$, $\pbarp$, and $\epem$):
\begin{enumerate}
\item The total number of primary 
charged particles produced in A+A reactions
($\nchaa$) scales linearly with the number of participants ($\np$).
\item The multiplicity per participant pair in A+A ($\nchaa/\halfnp$)
agrees with that measured in $\epem$ ($\nchee$) within 10\% over
a large range in $\s$ and $\snn$.  In general, the 
$\epem$ data has not fully removed
the contribution from weak decays, 
but this is generally less than a 10\% correction.
\item This agreement is not simply in the total multiplicity, but
extends over the full rapidity range (relative to the thrust
axis in the $\epem$ case).
\end{enumerate}
This agreement in total multiplicity suggests a certain 
universality in particle production, a result  consistent with
the thermal-statistical approach started by Fermi, Landau
and Hagedorn~\cite{Fermi:1950jd,Landau:gs,hagedorn}.  
However, the same data set
shows a systematic deviation between the multiplicities
below $\snn=30$ GeV, and increasing as the energy
gets lower.  It is the purpose of this letter to explore a simple
physics hypothesis which can explain this deviation in a semi-quantitative fashion. 
To compare multiplicities between
$\epem$ and heavy ion collisions we introduce the quantity  
\begin{equation}
\Delta \nch \equiv N_{e^+e^-} - \frac{N_{AA}}{\np/2}
\end{equation}
where, in an obvious notation, $N_{part}$ denotes the number of participants
in A+A collisions, while $N_{e^+e^-}$ is the total multiplicity 
in an $\epem$ collision, and $N_{AA}$ is the total multiplicity in an A+A
collision.
It is proposed that this difference is 
directly related to the thermodynamic variables determined
from the total multiplicities measured in A+A collisions, namely
the freeze-out values of the 
 temperature, $T$, and baryon chemical potential $\mu_B$.
For the multiplicity of charged particles  this relation reduces to
\begin{equation}
\Delta \nch = \frac{\mu_B}{3T}
\end{equation}
Various improvements on this relation will be discussed below.
 
Examination of the yields of different particle species over
a wide range in $\snn$ shows a large variation of the baryon
to pion ratio ($p/\pi$).  
Conversely, $p+p$ collisions show
a somewhat more rapid dependence of this ratio with beam energy.
This can be understood heuristically by saying that
relative to proton-proton collisions, collisions involving nuclei
(p+A, A+A) are distinguished by the larger ``stopping power'' of
the nuclear targets.  While this phenomenon is not well-understood
theoretically, it has been characterized phenomenologically
in several different ways.  
In proton-nucleus collisions, the
concept of ``rapidity loss'' is usually used to measure how
much energy the proton projectile loses in the multiple collisions
in the nuclear target~\cite{busza}.
While the data are not trivial to interpret, given the state of
understanding of longitudinal dynamics, one generally finds a
substantial net-baryon density near mid-rapidity in A+A collisions,
higher than p+p and $\epem$ reactions.

In Au+Au collisions, it is often postulated that 
the large particle multiplicities create a thermally and 
chemically equilibrated system.  This suggests using statistical
models to characterize the relative population of hadronic
states~\cite{Cleymans:2002mp,Braun-Munzinger:2003zd}.  In these
models, the main parameters are the freeze-out temperature ($T$),
which is associated with the energy density of the system,
and the baryochemical potential ($\mu_B$), which is directly
related to the density of the net baryon number 
distributed in the freeze-out volume ($N(p) - N(\overline{p}) 
\propto e^{\mu_B/T}-e^{-\mu_B/T}$).
It is less clear why similar fits should work for elementary
collisions (e.g. $\pp$ and $\epem$).
However, the work of Becattini has shown that statistical models 
prove to be
an equally useful tool in describing the relative yields of
hadrons in collisions with relatively small
multiplicities~\cite{Becattini:1997rv}, although additional
care must be taken to guarantee appropriate conservation of
quantum numbers (e.g. strangeness and baryon number).

\begin{figure*}[t]
\begin{center}
\includegraphics[width=15cm]{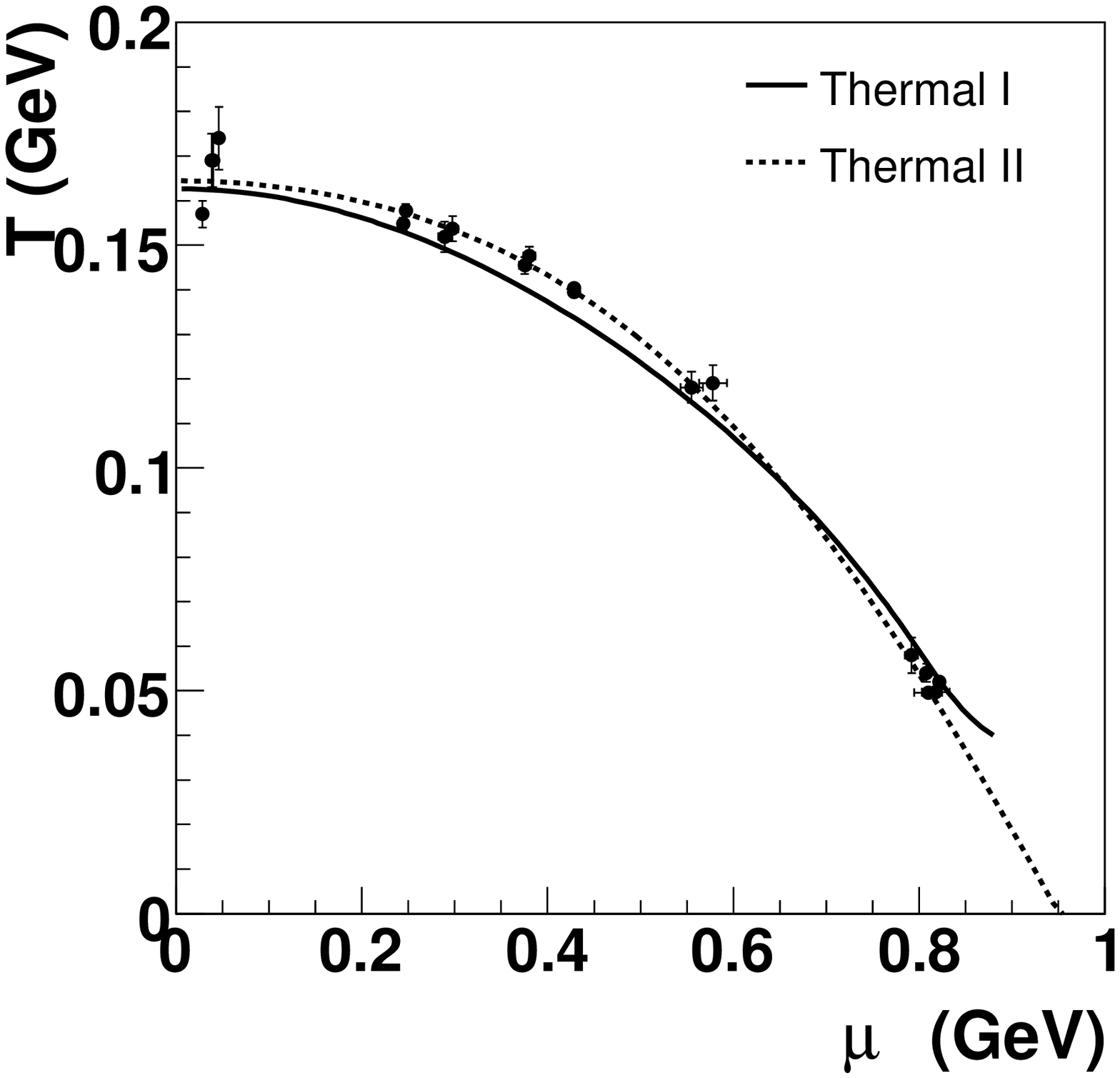}
\end{center}
\caption{
Compilation of thermal-statistical model fits to heavy ion
data over a wide range of beam energies.  Two parameterizations
of this data are shown, one based on $E/N\sim 1$ (solid line)
and the other a sixth-order polynomial fit (dotted line).
}
\label{tmub_2p}
\end{figure*}

Thermal fits made by a number of authors~\cite{Cleymans:2002mp,Braun-Munzinger:2003zd}
show that increasing the $\snn$ in A+A collisions leads
to an increase in $T$ and a correlated decrease in $\mu_B$,
shown in Fig. \ref{tmub_2p}.
This has been interpreted by Cleymans and Redlich by postulating a 
fixed relationship of the freezeout parameters, such that 
$\langle E \rangle / \langle N \rangle \sim 1$ GeV~\cite{Cleymans:1998fq}.  
Whatever the physical scenario implied by this condition, it provides a 
useful way to determine these parameters as a function of beam energy,
and to interpolate between available data points.
However, it turns out that this criterion (called ``Thermal I'') does
not perfectly describe the existing data.  A somewhat better description,
although purely phenomenological, can be made by 
a sixth-order polynomial fit in $\mu_B$
to the same data in the ($T$,$\mu_B$) plane
(``Thermal II'')~\cite{spencer}:
\[
T(\mu_B)=0.16446-0.11196\mu^2_B - 0.139139 \mu^4_B +0.0684637\mu^6_B
\]
In this work, we will show both parametrizations
where possible.

Also in this work, we use a parametrization of $\mu_B$ as a function
of $\s$ made by the authors in Ref. \cite{Cleymans:2004hj}
\begin{equation}
\mu_B(\s)=\frac{1.2735}{(1+0.2576\s)}
\end{equation}
To apply this information to the heavy ion and $\epem$ data,
we will invoke a simple thermodynamic condition.
When dealing with blackbody radiation, one typically sets the 
Gibbs potential $G=E-TS+pV = \sum_{i}\mu_i N_i \sim \mu_B N_B$, 
since the other chemical potentials (e.g. strangeness, charge,
isospin) are usually smaller than the baryochemical potential.
In this formula, $E$ is the internal energy, $T$ is the temperature, 
$S$ the entropy, $p$ the pressure, $\mu_B$ the baryochemical potential and 
$N_{B}$ the baryon number which must be conserved in the interaction.
This expression can be
rearranged to show how the entropy is related to the other variables:
\begin{equation}
S = \frac{(E+pV)-\mu_{B} N_{B}}{T} = S_0-S_B
\end{equation}
where
\begin{equation}
S_0=\frac{E+pV}{T}
\end{equation}
is the 
entropy due to the internal energy and the pressure of the system,
while
\begin{equation}
S_B=\frac{\mu_B N_{B}}{T}
\end{equation}
is interpreted as the entropy bound up in the conserved baryons, suppressing
the total entropy.

The $(E+pV)/T$ term can be understood 
as the one that controls particle production in
the absence of conserved baryon charges (i.e. $\mu_B=0$).  
It is assumed that this is universal for all strongly interacting 
collision systems with the same expansion features, most importantly
the dominance of 1D expansion in the early stages.  
The second term is thus a correction which will only be important
when $\mu_B N_{B}/T$ is non-negligible, i.e. at large $\mu_B$ or
small $T$ or both.  

This correction to the total entropy 
can be estimated in a crude way as follows:
\begin{itemize}
\item A factor of $\alpha=4$ to normalize entropy to the number of particles
(as is relevant for a massless Boltzmann gas).
\item A factor of $\np/2$ to give the total change in multiplicity
per participant baryon pair.  This cancels the $N_B$ in the numerator
since it is precisely the number of participants which determines
the conserved baryon number.
\item A factor of $\beta=3/2$ which accounts for unmeasured neutral pions.
This is based on the assumption that we are 
calculating the entropy of the lighter pions that would have been produced
except for the non-zero $\mu_B$ enforcing the presence of heavy baryons.
\end{itemize}
		
In other words, this scenario postulates $S/\nch=6$.  
Dividing by all these factors gives:
\begin{eqnarray}
\Delta \nch &=& \frac{2}{\alpha\beta\np}\frac{\mu_B N_B}{T}\\
\nonumber & = & \frac{2}{\alpha\beta} \times \frac{N_B}{\np} \times \frac{\mu_B}{T}\\
\nonumber & = & \frac{\mu_B}{3T}
\end{eqnarray}
This is the {\it multiplicity} 
that must be added to the low-energy results
at a given $\s$ to account for the entropy that would have been available
except for the need to conserve baryon charge.  

It turns out that direct calculations with thermal 
models~\cite{thermus}, give the result that the entropy
divided by the multiplicity of final-state charged particles (after strong
decays) is $S/\nch=7.2$.  This number should be compared to $\alpha\beta=6$ from
the considerations above.  For the subsequent calculations, the more
theoretically relevant number will be used instead of the simpler estimate.
The difference between them should be seen as contributing to an overall
theoretical uncertainty.

\begin{figure*}[t]
\begin{center}
\includegraphics[width=15cm]{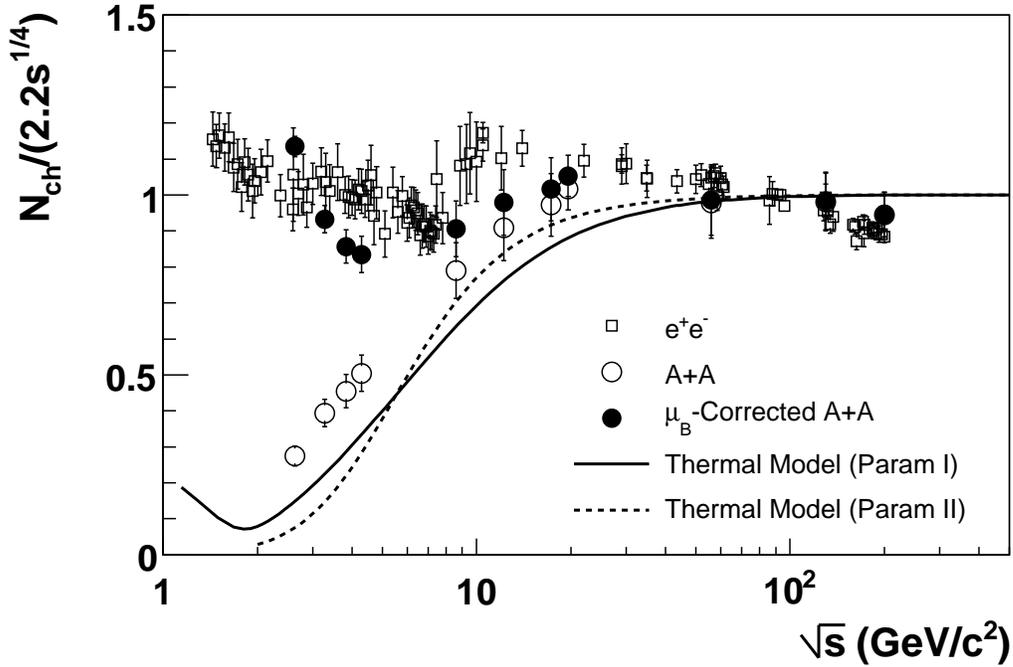}
\end{center}
\caption{
Ratio of $\nch$ for $\epem$ annihilation (open squares)
and $\nch/\halfnp$ for A+A collisions (open circles) to
a Landau-hydro expression for the total multiplicity ($2.2s^{1/4}$).
Also shown are the same ratios for A+A after correcting by 
$\mu_{B}/(\alpha\beta/2)T$ (closed circles).
The ratio of the entropy density to the asymptotic value as a function
of $\sqrt{s}(\mu_B)$ is shown
for the two paths through the $T,\mu_B$ plane shown in Fig. 1.
These should
be compared with the open circles, as it already contains the effects 
described in the main text.
}
\label{ee_AA_mub_thermal}
\end{figure*}

When this is done, we get the results shown in Fig.~\ref{ee_AA_mub_thermal},
where the multiplicities have been divided by the Landau-Fermi
expression ($N_{ch}=2.2s^{1/4}$)~\cite{Fermi:1950jd,Landau:gs,Afanasiev:2002mx}.
The $\epem$ results are shown as open squares,
the original A+A results are shown as open circles, and
the ``$\mu_{B}$''-corrected A+A results are shown as closed circles.
It is surprising that this simple model works as well as it does,
since it is nothing more than correcting for the fact that the
initial baryons must be present in the final state, and thus
take up energy that would have normally gone to normal thermal
particle production (mainly pions).

\begin{figure*}[t]
\begin{center}
\includegraphics[width=15cm]{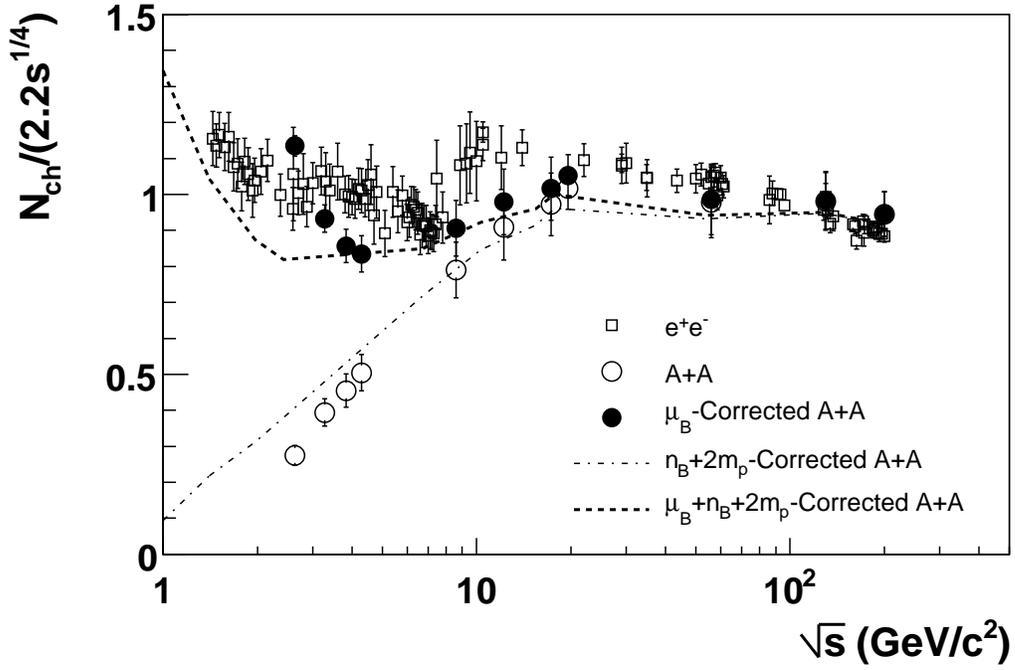}
\end{center}
\caption{
Ratio of $\nch$ for $\epem$ annihilation (open squares)
and $\nch/\halfnp$ for A+A collisions (open circles) to
a Landau-hydro expression for the total multiplicity ($2.2s^{1/4}$).
Also shown are the same ratios for A+A after various corrections:
1) $\mu_{B}/3T$, $n_B$, $2m_P$ corrections (dotted line)
and 2) $n_B$, $2m_P$ corrections (dot-dashed line).
}
\label{ee_AA_mub_thermal_nbmp}
\end{figure*}

However, it can be argued that the $4\pi$ multiplicity
collisions
has an extra component that would not be found in $\epem$ collisions,
namely the participant baryons themselves.  To correct for these, we 
make two additional transformations on the A+A data.
\begin{itemize}
\item ``$m_P$'': Subtract $2m_P$ from $\s$ to 
correct for  the difference in the  mass of the beam particle
\item ``$n_B$'': Subtract 1, assuming that the net baryons will be
either $p$ or $n$ equally.
\end{itemize}
These corrections are shown to have relatively little effect on
the basic result, as seen by the thick dotted line in Fig.~\ref{ee_AA_mub_thermal_nbmp}.
Thus, it is not possible to distinguish their
relevance by comparison with the $\epem$ data.
It should also be noted that just applying the $m_P$ and $n_B$ corrections,
without the $\mu_B$ correction, has little effect on the initial result
as well, as shown in Fig.~\ref{ee_AA_mub_thermal_nbmp}.


A more quantitative check of the physical picture discussed 
here can be made by  calculating
the entropy density vs. $\mu_B$ in a full statistical-thermal 
model calculation~\cite{thermus,Maciej}.
The freezeout contour is the same as in the calculation above.
To relate this curve to experimental data, one considers the translation
of the experimental variable $N_{ch}^{A+A}/(N_{ch}^{\epem}\np/2)$ to 
parameters accessible in thermal models.
From the fits performed on $p+p$ and $\epem$ 
data in Ref.~\cite{Becattini:1997rv}, we find that the $T$ parameter is essentially
constant over a wide range of energy $T=T_0 \sim 170$ MeV.
In this case, $s(T,\mu_B) = s(T_0,0)$ and thus should be constant with $\s$.
It also appears to be the asymptotic value reached in ultra-high energy 
$A+A$ collisions, if current trends are to be believed.  
In this case, we assume that 
\begin{equation}
N^{\epem}_{ch} = C^{\epem} V^{\epem} s_0
\end{equation} 
where
$V^{\epem}$ is the freezeout volume and $C_{\epem}$ is a constant relating
the number of detected particles to the total entropy (again, approximately
4).  In this picture, the energy dependence of the 
total multiplicity is determined dominantly by the freezeout volume.

In A+A collisions, the multiplicity has been found to scale linearly
with the number of participating nucleons $\np$.  Since the volume of the
initial nuclei scales with $A$, and the freezeout entropy scales with $V$,
this suggests that $V \propto \np/2 = V^{A+A} \np/2$.
We then assume that  
\begin{equation}
N^{A+A}_{ch} = \frac{\np}{2} C^{A+A} V^{A+A} s(T,\mu_B)
\end{equation}
where $V_{A+A}$ is the effective volume {\it per participant pair}.

With these assumptions, the ratio shown in Figure 3 is
\begin{equation}
\frac{2}{\np}
\frac{N_{ch}^{A+A}}{N_{ch}^{\epem} } = 
\frac{C^{A+A}}{C^{\epem}} 
\frac{V^{A+A}}{V^{\epem}} 
\frac{s(T,\mu_B)}{s_0}.
\end{equation}

The right-hand side of this equation has two sets of constants (C and V)
and one ratio that depends on beam energy.
The constants C control the proportionality between the total entropy
and the total charged particle multiplicity.  
Landau and Belenkij~\cite{Belenkij:cd}
argued that this constant does not depend on the system size, so it makes sense
to set this ratio to unity.  This presumption is supported by the overall
similarity in the particle production~\cite{Becattini:1996gy} (although
strangeness is clearly suppressed in the smaller systems).
They also did not think there would be a change in the relation of the multiplicity
to the entropy as a function of beam energy or initial baryon density.
The ratio of the volumes might not be expected to be the same.  However, 
the agreement (to the 10\% level) of the total multiplicity per participant
pair in A+A and the total multiplicity in $\epem$ suggests that these
volumes are similar, even as a function of energy.

In Fig.~\ref{ee_AA_mub_thermal} we compare the ratio $s(T,\mu_B)/s_0$, shown 
for the two parametrizations of $T(\mu_B)$ (Thermal I and Thermal II),
with the ratios discussed previously.
The agreement between the thermal model curve and the A+A 
data is in
reasonable qualitative agreement.  This suggests that the previously-made
physics assumption
\begin{equation}
\frac{C^{A+A}}{C^{\epem}}\frac{V^{A+A}}{V^{\epem}}=1
\end{equation}
is a reasonable one.

It should be noted that the concept of ``pion suppression'' was previously
discussed by Gazdzicki et al~\cite{Gazdzicki:1997sg} 
by reference to data from a similar set
of experiments as discussed in~\cite{Back:2003xk}.  
However, the phenomenon discussed in this work is concerned with 
the suppression of the total
entropy rather than just the pions, and is a feature which arises naturally
in the context of thermal models.  The models here also include all available
meson and baryon resonances, as opposed to just delta resonances.  

\begin{figure*}[t]
\begin{center}
\includegraphics[width=15cm]{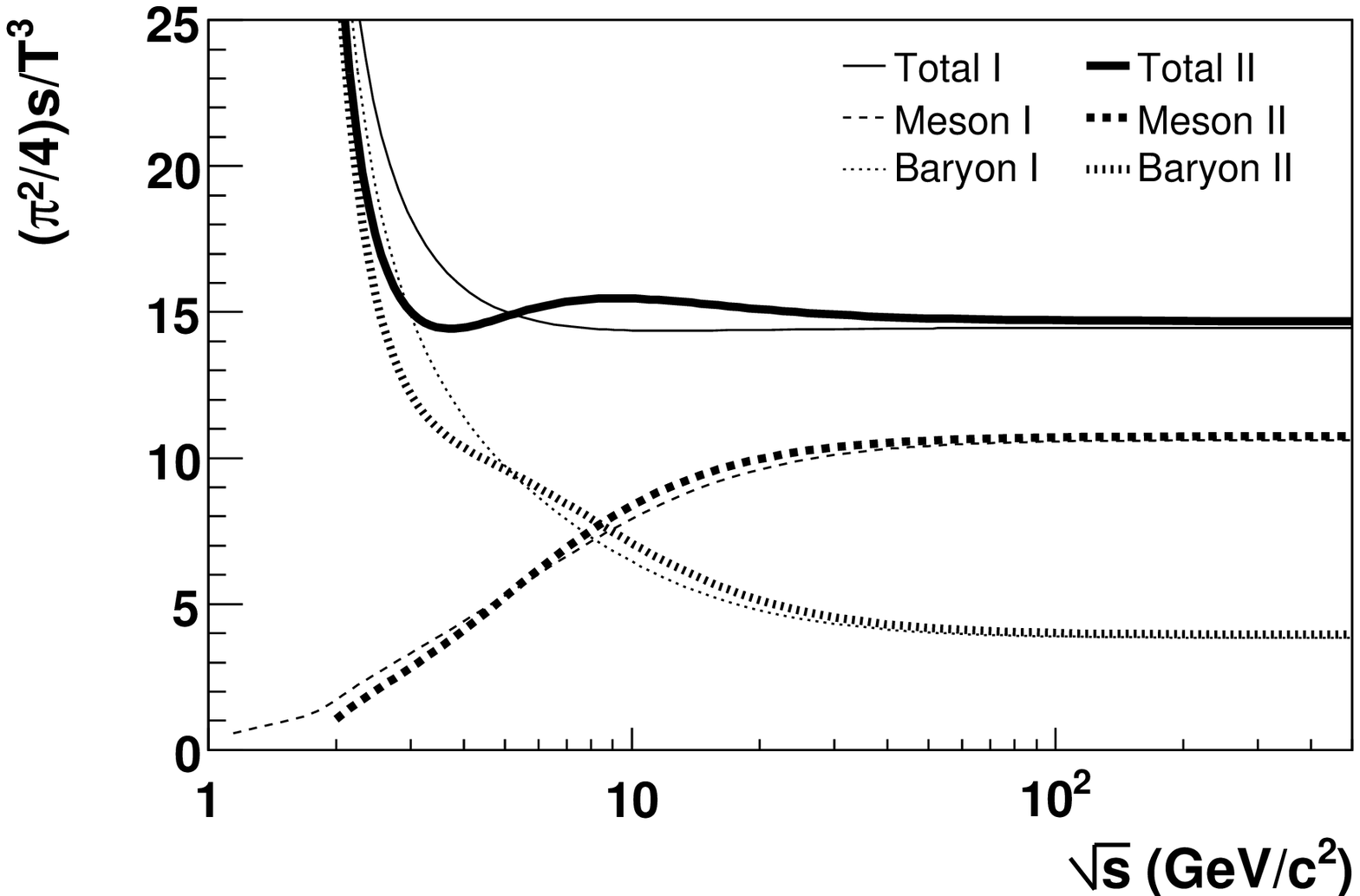}
\end{center}
\caption{
Entropy density normalized by $T^3$ and scaled by $\pi^2/4$,
shown separately for mesons
and baryons as a function of $\sqrt{s}(\mu_B)$.  
For each species, the two parametrizations of $T(\mu_B)$ are
shown, giving an estimate of the systematic uncertainty.
While the two components exchange their relatve dominance as a function
of beam energy, their total is constant over most of the
calculated energies.  
}
\label{entropy_sqrts}
\end{figure*}

To make the relative energy dependence of mesons and baryons clearer, we show
their respective contributions to the entropy density as a function of beam
energy in Fig.~\ref{entropy_sqrts}, again for two parametrizations.
The individual and total entropy densities are both divided
by $T^3$ to remove the expected temperature dependence.  
They are then multiplied
by a factor if $\pi^2/4$, which transforms the quantity $s/T^3$ into
the effective number of degrees of freedom $n(T)$ of a massless Boltzmann gas.
The main result is that the baryon contribution completely dominates
at low energies, but the mesons are equal at $\snn\sim 10$ GeV and
their contribution exceeds that of the baryons by a factor of $\sim 2$
and saturates.  
However, as was also noted in Ref.~\cite{Tawfik:2004ss},
it is observed that the quantity $s/T^3$ is constant over a large
range in center of mass energies (down to $\snn=5$ GeV) and 
diverges only at very low energies (presumably due to the associated
rapid decrease in $T$).  
This result
shows that not only are the number of degrees of freedom similar at freezeout
for $A+A$ and $\epem$, but they are similar for heavy ion collisions over
a large range of collision energies.  
From this result, that freezeout occurs for $s/T^3=const.$,
one might also explain the suppression of the entropy density
$s=const.\times T^3$,
as due to the lower temperatures associated with larger $\mu_B$.
In either case, the suppression of the total multiplicity results  from the
non-trivial interplay between $\mu_B$ and $T$.

In conclusion, the difference between the
charged particle multiplicity per participant pair in A+A 
and the multiplicity in $\epem$ can be explained by the suppression
of entropy due to the presence of a conserved quantum number,
manifest as the net-baryon density.
A semi-quantitative understanding of the existing data has been achieved both
by simple thermodynamic arguments as well as more detailed
thermal model comparisons.

This work was partially supported by the US DOE grant DE-AC02-98CH10886.
P.S. acknowledges fruitful discussions with 
Mark Baker,
Francesco Becattini,
Nigel George,
Dan Magestro
and
Gunther Roland.

\begin{thebibliography}{50}

\bibitem{Back:2003xk} B.~B.~Back {\it et al.}, arXiv:nucl-ex/0301017.
\bibitem{Fermi:1950jd} E.~Fermi, Prog.\ Theor.\ Phys.\  {\bf 5}, 570 (1950).
\bibitem{Landau:gs} L.~D.~Landau, Izv.\ Akad.\ Nauk Ser.\ Fiz.\  {\bf 17}, 51 (1953).
\bibitem{hagedorn} R.~Hagedorn, CERN-Report 71-12 (1971).
\bibitem{busza} W.~Busza and A.~S.~Goldhaber, Phys.\ Lett.\ B {\bf 139}, 235 (1984).
\bibitem{Cleymans:2002mp} J.~Cleymans, arXiv:hep-ph/0201142.
\bibitem{Braun-Munzinger:2003zd} P.~Braun-Munzinger, K.~Redlich and J.~Stachel, arXiv:nucl-th/0304013.
\bibitem{Becattini:1997rv}
F.~Becattini and U.~W.~Heinz, Z.\ Phys.\ C {\bf 76}, 269 (1997) [Erratum-ibid.\ C {\bf 76}, 578 (1997)].
\bibitem{Cleymans:1998fq} J.~Cleymans and K.~Redlich, Phys.\ Rev.\ Lett.\  {\bf 81}, 5284 (1998).
\bibitem{spencer} S.~Wheaton, PhD. Thesis from University of Cape Town (2005).
\bibitem{Cleymans:2004hj} J.~Cleymans, H.~Oeschler, K.~Redlich and S.~Wheaton, arXiv:hep-ph/0411187.
\bibitem{thermus} S.~Wheaton and J.~Cleymans, arXiv:hep-ph/0407174.
\bibitem{Afanasiev:2002mx} S.~V.~Afanasiev {\it et al.} ,Phys.\ Rev.\ C {\bf 66}, 054902 (2002).
\bibitem{Maciej} M. Stankiewicz, Honours Thesis, University of Cape Town (unpublished, 2004).
\bibitem{Belenkij:cd} S.~Z.~Belenkij and L.~D.~Landau, Nuovo Cim.\ Suppl.\  {\bf 3S10}, 15 (1956) [Usp.\ Fiz.\ Nauk {\bf 56}, 309 (1955)].
\bibitem{Becattini:1996gy} F.~Becattini, arXiv:hep-ph/9701275.
\bibitem{Gazdzicki:1997sg} M.~Gazdzicki, M.~I.~Gorenstein and S.~Mrowczynski, Eur.\ Phys.\ J.\ C {\bf 5}, 129 (1998).
\bibitem{Tawfik:2004ss} A.~Tawfik, arXiv:hep-ph/0410392.
\end{thebibliography}
\end{document}